\documentclass[11pt]{article}

\usepackage[latin2]{inputenc}

\usepackage[english]{babel}

\usepackage{multirow}

\usepackage{graphicx}
\usepackage{psfrag}

\usepackage{color}
\usepackage{amsfonts}
\usepackage{amssymb}
\usepackage{hyperref}

\title{On messengers couplings in extended GMSB models}
\date{}

\author{T. Jeli\'{n}ski
\medskip
\\{\small {\it 
Department of Field Theory and Particle Physics, 
}}\\
{\small {\it Institute of Physics, University of Silesia,}}\\{\small {\it 
Uniwersytecka 4, 
40-007 Katowice, Poland}}\\}

\def\beq{\begin{equation}}
\def\eeq{\end{equation}}

\def\beqa{\begin{eqnarray}}
\def\eeqa{\end{eqnarray}}

\def\bar{\begin{array}}
\def\ear{\end{array}}

\def\ben{\begin{enumerate}}
\def\een{\end{enumerate}}

\def\bit{\begin{itemize}}
\def\eit{\end{itemize}}

\def\a{\alpha}

\def\t{\theta}

\def\la{\lambda}

\def\jd{\frac{1}{2}}

\newcommand{\eqref}[1]{(\ref{#1})}

\def\nn{\nonumber}

\def\phifb{\phi_{\overline{5}}}
\def\phit{\phi_{10}}
\def\Yt{Y_{10}}
\def\Ytb{Y_{\overline{10}}}
\def\Yf{Y_{5}}
\def\Yfb{Y_{\overline{5}}}
\def\Hf{H_5}
\def\Hfb{H_{\overline{5}}}

\def\sY{\mathsf{Y}}

\begin{document}

\maketitle

\abstract{
We discuss, for the first time, the role of the superpotential couplings of three messenger fields in a GMSB-type unification model in which messenger sector consists of both $5+\overline{5}$ and $10+\overline{10}$ of $SU(5)$. It turns out that these interactions are relevant when coexist with appropriate messenger-MSSM couplings. Then they induce extra contributions to 2-loop soft masses. In the present model, we derive all such soft terms and discuss constraints which have to be satisfied to avoid rapid proton decay and $\mu/B_{\mu}$ problem. As an example, it is shown how superpotential couplings
of three messengers influence mass spectra when the model is restricted by additional global $U(1)_{q}$ symmetry. We find that 
masses of the lightest sleptons are the most sensitive to those new interactions, what in some cases results in the NLSP/NNLSP pattern in which stau or selectron is lighter than the lightest neutralino. 
}

\section{Introduction} 

\normalsize
\noindent 
The recent measurement of the Higgs-like boson mass at the LHC \cite{Aad:2012tfa,Chatrchyan:2012ufa} triggers a lot of questions about its consequences for models of SUSY breaking mediation \cite{Arbey:2011ab}. It is well-known that $m_{h^{0}}\approx125$ GeV can be achieved in the MSSM by adequate left-right stops mixing, which in turn originates e.g. from large $A$-terms at the EWSB scale \cite{Draper:2011aa}. It is hard to accommodate them in the standard Gauge-Mediated Supersymmetry Breaking (GMSB) models, but they naturally arise at 1-loop level in so-called extended\footnote{Models which include marginal couplings of messengers are also called  Yukawa-Deflected, More Generic or Flavoured Gauge Mediation models.} GMSB models \cite{Dine:1996xk,Giudice:1997ni,Chacko:2001km,Evans:2013kxa}. In that class of frameworks, messengers interact with themselves and with MSSM matter fields via renormalizable superpotential couplings (herein called marginal couplings).\footnote{We adopt terminology used in \cite{Evans:2013kxa}.} One usually  considers two types of them: matter-matter-messenger and matter-messenger-messenger. That topic has been widely investigated from quite a long time in various settings, and it is well known that some of those couplings significantly change phenomenology of the usual GMSB models \cite{Giudice:1998bp} because they generate additional 1- and 2-loop soft terms. Depending on the context, the focus was on messenger-Higgs \cite{Kang:2012ra,Craig:2012xp}, messenger-quark \cite{Evans:2011be} or messenger-lepton \cite{Shadmi:2011hs} interactions. Selection rules are usually delivered by R-parity, or some global $U(1)$ symmetry, which in some cases is related to Froggatt-Nielsen mechanism. For simplicity, it is often assumed that only couplings to the third family of the matter are relevant (i.e. some hierarchy of interactions is assumed, akin to the one in MSSM Yukawas). However, there are attempts to justify the structure of the couplings and address the issue of FCNC in the case when messengers interact also with the light families of MSSM \cite{Shadmi:2011hs,Albaid:2012qk,Abdullah:2012tq,Calibbi:2013mka}. The upshot being that relatively small hierarchy (much milder than the one in MSSM Yukawas) renders predictions of those models perfectly consistent with low-energy observables. Recently, all matter-matter-messenger and matter-messenger-messenger couplings have been analysed in the case when messengers are in $SU(5)$ representations of low dimension (singlet, fundamental, antisymmetric and adjoint) \cite{Evans:2013kxa,Byakti:2013ti}. Moreover, the wave-function renormalization method, which is relevant for computing soft terms, was substantially improved \cite{Evans:2013kxa}. Finally, it is worthwhile to mention that the attractive feature of extended GMSB models is that they not only accommodate for large $A$-terms, but also in some cases realize non-standard NLSP/NNLSP mass patterns (e.g. stop/bino) with NLSP mass in the range that may be probed at the LHC \cite{Kats:2011it}.

In this letter we analyse the possible marginal couplings of messengers in GMSB-type model which contains one pair of messengers in representation $5+\overline{5}$ and one in $10+\overline{10}$ of $SU(5)$. The  novel elements are superpotential couplings of the form: messenger-messenger-messenger. We show that they  are relevant for the phenomenology when coexist with appropriate messenger-messenger-matter or messenger-matter-matter interactions. Then they contribute to 2-loop soft masses. At the same time, they do not generate  
$A$-terms nor give contributions to 1-loop soft masses. 
The latter feature is expected to be important for low-scale GMSB models.

One can check that proper phenomenology put several constraints on the discussed couplings. Some of them must be highly suppressed to avoid inducing operators which would lead to rapid proton decay, or those which make it difficult to realize EWSB \cite{Dvali:1996cu}. In the model under consideration, we analyse the issue of baryon/lepton number violation and generation of $\mu$ and $B_{\mu}$ terms at 1-loop. To elude mentioned problems extra global $U(1)_{q}$ symmetry is introduced. Detailed analysis of the phenomenology of the models restricted by that symmetry shows that, even in such simplified frameworks, superpotential couplings 
of three messengers influence  mass spectrum significantly (e.g. by changing NLSP/NNLSP pattern).

The paper is organized as follows. In Section 2 we recall details of the extended GMSB model, and analyse soft terms generated by marginal couplings of messengers. Section \ref{softeta} contains new results. Namely, we derive 2-loop contributions to soft masses induced by superpotential couplings of three messengers. 
In Section \ref{U(1)} it is shown how to avoid the proton decay and $\mu/B_{\mu}$ problem by invoking additional global $U(1)_{q}$ symmetry. In Section 3 we give examples of the simplest models restricted by that symmetry, and investigate their low-energy phenomenology.  We conclude in Section 4. Appendices contain tables of $U(1)_{q}$ charges and numerical coefficients related to 2-loop soft masses generated by messenger-messenger-messenger couplings.

\section{Extended GMSB model}

\noindent We consider $SU(5)$ unification model of GMSB-type in which SUSY breaking effects are communicated to the visible sector through two pairs of messengers: $(\Yf,\Yfb)$ and $(\Yt,\Ytb)$, where subscripts denote representations of $SU(5)$. 
In the visible sector all matter fields, beside Higgses, can be arranged into $\overline{5}$ and $10$ of $SU(5)$.\footnote{Higgs fields are denoted by $\Hf$ and $\Hfb$, while $\phifb$ and $\phit$ stand for the superfields containing quarks $Q, U, D$ and leptons $L,E$ of the third family and their superpartners. Subscripts indicate $SU(5)$ representations. All matter superfields are collectively denoted by $\Phi$. Abusing notation, we denote matter superfields with the same symbol as their Standard Model components.}
Higgs triplets are supposed to have mass of order GUT scale $M_{GUT}$, and in the theory below $M_{GUT}$ they are absent  from the spectrum. However, it is very convenient to use $SU(5)$ notation though - all couplings involving Higgs fields $H_{5,\overline{5}}$ are understood to be projected onto doublet components of $5,\overline{5}$. 

We assume that the SUSY breaking in the hidden sector can be 
parametrized by gauge singlet chiral superfield $X$ (so-called spurion) which
lowest component and $F$-term spontaneously get vev i.e. 
$\left<X\right>=M+\theta^2F$. As usual, the following superpotential couplings between spurion and messengers
\beq\label{WX}
W_{X}=X(Y_{5}Y_{\overline{5}}+Y_{10}Y_{\overline{10}})
\eeq
provide mass $M$ for the latter. We choose $M$ to be of order $10^{14}$ GeV. 
For the simplicity, it is assumed that in \eqref{WX} the spurion  coupling to both pairs of messengers is the same. SUSY breaking effects are transmitted from the hidden sector to MSSM via the messengers. 
In the extended Gauge Mediation models, messengers interactions which are relevant for the mediation mechanism are not only couplings to gauge fields and to the spurion, but also to MSSM fields. Therefore, we consider all marginal superpotential couplings of messengers and MSSM matter that are allowed by gauge symmetry (including couplings of three messengers). They can be organized in terms of $SU(5)$ invariants. It is easy to check that in the model under consideration such terms are of the form:\footnote{ 
All such terms which involve messengers can a priori appear while in the visible sector one needs only couplings: $\Hf\phit\phit$ and $\Hfb\phifb\phit$. The issue of dangerous operator $\phifb\phifb\phit$ 
shall be discussed in the Section \ref{dangerous}.} $5\,10\,10$, $\overline{5}\,\overline{10}\,\overline{10}$, $5\,5\,\overline{10}$ and $\overline{5}\,\overline{5}\,10$. 
We assume that they are hierarchical i.e. only coupling to the heaviest family of the MSSM is of order one 
- the interactions with other two families are assumed to be small enough not to induce large FCNC effects.     
Taking into account full flavour structure of the Yukawas and messenger couplings to the first and second family is left for the future work.

\subsection{Marginal couplings of messengers and MSSM matter}\label{softh}

The part of the superpotential which contains marginal couplings of messengers $\sY=\{\Yf,\Yfb,\Yt,\Ytb\}$ is of the following form
\beq\label{WY}
W_{\sY}=W_{\sY\sY\sY}+W_{\Phi\Phi\sY}+W_{\Phi\sY\sY},
\eeq
where $W_{\sY\sY\sY}$ are novel interactions of three messengers. They are crucial for the further discussion, and they will be discussed in details in the next section. On the other hand, the other two terms in \eqref{WY} include messenger couplings to matter fields $\Phi=\{\Hf,\Hfb,\phifb,\phit\}$ i.e.\footnote{For the simplicity, we assume that all superpotential coupling constants are real.}
\beq\label{WII}
W_{\Phi\Phi\sY}=h_1\Hf\phit\Yt+\jd h_2\phit\phit\Yf+h_3\Hfb\phifb\Yt+h_4\Hfb\phit\Yfb+\jd h_5\phifb\phifb\Yt+h_6\phifb\phit\Yfb
\eeq
are matter-matter-messenger couplings, and
\beqa\label{WI}
W_{\Phi\sY\sY}&=&\frac{1}{2}h_7\Hf\Yt\Yt+h_8\phit\Yf\Yt+\jd h_9\Hfb\Ytb\Ytb+\jd h_{10}\phifb\Ytb\Ytb\\
&&+h_{11}\Hf\Yf\Ytb+h_{12}\Hfb\Yfb\Yt+h_{13}\phifb\Yfb\Yt+\jd h_{14}\phit\Yfb\Yfb\nn
\eeqa
are matter-messenger-messenger couplings. We shall denote coupling constants $h_A$ collectively by $h$.  
Let us recall that in the extended GMSB models one usually considers only interactions of messengers $\sY$ and MSSM matter fields $\Phi$ of those two types listed above: \eqref{WII} and \eqref{WI}.\footnote{$W_{\Phi\Phi\sY}$ and $W_{\Phi\sY\sY}$ are sometimes called, respectively, type II and type I messenger couplings \cite{Evans:2013kxa}.} Both of them, in fact, change predictions of the standard GMSB models in several ways e.g. by enhancing left-right squarks mixing, or 
allowing various non-standard types of NLSP/NNLSP patterns \cite{Jelinski:2011xe}. 

Below messenger scale $M$, one gets MSSM with soft terms. Gaugino masses $M_{\lambda}^{(r)}$ are of the same form as in the GMSB models i.e. they arise at 1-loop
\beq
M^{(r)}_{\la}=\frac{\a_r}{4\pi}n_X\xi \, ,
\eeq
where $r=1,2,3$ corresponds to the gauge group $U(1)_Y$, $SU(2)_L$ and $SU(3)_C$ of the
Standard Model (we use the GUT normalization for the hypercharge), and $n_X=4$ is twice the sum of the Dynkin indices of the messenger fields coupling to the spurion $X$. $\xi=F/M$ is the scale of gauginos and scalar soft masses. We choose $F\sim10^{19}\,\textrm{GeV}^2$ hence $\xi\sim10^5$ GeV.\footnote{In the discussed model gravitino is the LSP with mass $m_{3/2}\sim1$ GeV.}  Masses $M_{\lambda}^{(r)}$ do not depend on marginal messenger couplings at the leading order. On the other hand, $h_{A}$ do contribute to both 1-loop $A$-terms and 2-loop soft masses.\footnote{The 1-loop soft masses generated by $h_{A}$ are negligible because of $\xi/M\ll1$ .} 
They can be derived with the help of wave-function renormalization method \cite{Giudice:1997ni,Chacko:2001km,Evans:2013kxa}. One can find that 
$A$-terms generated 
by \eqref{WII} and \eqref{WI} are of the form\footnote{When messengers couplings to the first and second family of the MSSM matter are relevant then also R-parity violating $A$-terms $L^iL^jE^k$ and $U^iU^jD^k$
can appear.}
\beqa
\label{Aterms}
A_{t,b,\tau}=-\frac{\xi}{4\pi}\sum_{A}C^{(t,b,\tau)}_{A}\a_{h_{A}}\quad\textrm{and}\quad A_b'=-\frac{\xi}{4\pi}\sum_{A<B}C'^{(b)}_{A,B}(\alpha_{h_A}\alpha_{h_B})^{1/2}.
\eeqa
The numerical coefficients $C^{(t,b,\tau)}_{A}$ and $C'^{(b)}_{A,B}$ are given in the Table \ref{tabAcoeff}. The $A$-terms given in \eqref{Aterms} are related to the trilinear terms in the scalar potential $V$ in the following way:
\beq
V\supset y_{t}A_{t}H_{u}\widetilde{Q}\widetilde{U}+y_{b}A_{b}H_{d}\widetilde{Q}\widetilde{D}+y_{\tau}A_{\tau}H_{d}\widetilde{L}\widetilde{E}+y_{b}A'_{b}\widetilde{L}\widetilde{Q}\widetilde{D},
\eeq
where $y_{t,b,\tau}$ denote MSSM Yukawa couplings of the third family. The scalars $\widetilde{\Phi}\in\{H_u,H_d,\widetilde{L},\widetilde{E},\widetilde{Q},\widetilde{U},\widetilde{D}\}$ receive 2-loop corrections to soft SUSY breaking mass terms from three sources what can be written as
\beq
\label{m2soft}
m_{\widetilde{\Phi}}^2=m_{\widetilde{\Phi},g}^2+m_{\widetilde{\Phi},h}^2 + m_{\widetilde{\Phi},\eta}^2.
\eeq
$m^2_{\widetilde{\Phi},g}$ are well-known 2-loop 
gauge mediation mass terms induced by gauge interactions transmitting SUSY breaking from messenger sector \cite{Giudice:1998bp}
\beqa
m_{\widetilde{\Phi},g}^2&=&2\sum\limits_{r=1}^3C_2^r(\Phi)\frac{\a_r^2}{(4\pi)^2}n_X\xi^2,
\eeqa
where $C_2^r(\Phi)$ are quadratic Casimir operators of the representation of $\widetilde{\Phi}$ under $r$-th gauge group.
The contributions $m^2_{\widetilde{\Phi},h}$ in (\ref{m2soft}) are 2-loop terms induced by the superpotential couplings \eqref{WII} and \eqref{WI}. Since formulas for $m^2_{\widetilde{\Phi},h}$ are quite lengthy, and we do not need their explicit form here, we shall not list them at this point.\footnote{The general formulas for 2-loop masses induced by \eqref{WII}, \eqref{WI} and \eqref{WYYY} can be found in the ancillary Mathematica file \texttt{m2Phi.nb}.} Lastly, $m^2_{\widetilde{\Phi},\eta}$ are new contributions to soft masses of scalars $\widetilde{\Phi}$ which are generated by marginal couplings of three messengers $W_{\sY\sY\sY}$. They are discussed below.
\begin{table}
\hspace{15mm}\parbox{.45\linewidth}{
\begin{center}
\footnotesize 
\begin{tabular}{|c|c|c|c|c|c|c|c|c|c|c|c|c|c|c|}
\cline{2-15}
\multicolumn{1}{c|}{}&\multicolumn{14}{c|}{$A$}\\
\cline{2-15}
\multicolumn{1}{c|}{}& 1 & 2 & 3 & 4 & 5 & 6 & 7 & 8 & 9 & 10 & 11 & 12 & 13 & 14\\
\hline
$C^{(t)}_{A}$ & 9 & 6  & 0 & 1 & 0 & 4 & 3 & 6 & 0 & 0 & 4 & 0 & 0 & 1\\
$C^{(b)}_{A}$ & 1 & 3 & 6 & 5 & 2 & 6 & 0 & 3 & 3 & 3 & 0 & 4 & 4 & 1\\
$C^{(\tau)}_{A}$ & 0 & 3 & 5 & 6 & 3 & 6 & 0 & 3 & 3 & 3 & 0 & 4 & 4 & 0\\
\hline
\end{tabular}
\end{center}
}
\hfill
\hspace{5mm}
\parbox{.45\linewidth}{
\vspace{-8.5mm}
\begin{center}
\footnotesize
\begin{tabular}{|c|c|c|c|c|}
\cline{2-5}
\multicolumn{1}{c}{}&\multicolumn{4}{|c|}{$A,B$}\\
\cline{2-5}
\multicolumn{1}{c|}{}& 3,5 & 4,6 & 9,10 & 12,13\\
\hline
$C'^{(b)}_{A,B}$ & 3 & 4 & 3& 4\\
\hline
\end{tabular}
\end{center}
}
\caption{Numerical  coefficients $C^{(t,b,\tau)}_{A}$ and $C'^{(b)}_{A,B}$ in 
the $A$-terms \eqref{Aterms}.
}\label{tabAcoeff}
\end{table}
 
\subsection{Marginal couplings of three messengers}\label{softeta}

The part of the superpotential which contains interactions of three messenger fields can be written as
\beq\label{WYYY}
W_{\sY\sY\sY}=\jd\left(\eta_1\Yf\Yt\Yt+\eta_2\Yfb\Ytb\Ytb+\eta_3\Yf\Yf\Ytb+\eta_4\Yfb\Yfb\Yt\right),
\eeq
where $\eta_{i}$ are coupling constants of order one, collectively denoted by $\eta$.  In the discussed framework, \eqref{WYYY} induce extra corrections to scalar soft masses $m^{2}_{\widetilde{\Phi}}{\widetilde{\Phi}}^{\dag}\widetilde{\Phi}$. As in the case of $h$ couplings, one can obtain them using wave-function renormalization method \cite{Giudice:1997ni,Chacko:2001km,Evans:2013kxa}. Those new terms can be written in the following form
\beqa\label{m2eta} 
m_{\widetilde{\Phi},\eta}^2&=&\frac{\xi^2}{16\pi^2}\sum_{iA\leq B\leq Cf}\left(C^{(\Phi)}_{i,A}\a_{\eta_i}\a_{h_A}+C^{(\Phi)}_{i,(A,B,C)}(\a_{\eta_{i}}\a_{h_A}\a_{h_B}\a_{h_C})^{1/2}\right.\nn\\
&&\phantom{\frac{\xi^2}{16\pi^2}\sum_{ijA\leq B\leq C}(}\left.+C^{(\Phi)}_{i,(A,B,f)}(\a_{\eta_{i}}\a_{h_A}\a_{h_B}\a_{f})^{1/2}\right),
\eeqa
where $C^{(\Phi)}$'s are numerical constants while $\a_{h_{A}}=h_{A}^{2}/(4\pi)$, $\a_{\eta_{i}}=\eta_{i}^{2}/(4\pi)$. $h_A$ are defined in \eqref{WII} and \eqref{WI}. Finally, $\a_{f}$'s are related to MSSM Yukawa couplings $y_{t,b,\tau}$: $\a_{f}=y_{f}^{2}/(4\pi)$, $f=t,b,\tau$. 
Full list of coefficients $C^{(\Phi)}$ can be found in the Appendix \ref{appCs}. The 2-loop contributions to soft masses displayed in eq. \eqref{m2eta} are new results for the discussed class of extended GMSB models. The consecutive components of the sum \eqref{m2eta} arise from 2-loop diagrams with two $h$ and two $\eta$ vertices, three $h$ and one $\eta$ vertex and from diagram with two $h$, one $\eta$ and one $y_{f}$ vertex respectively. Let us mention that beside \eqref{m2eta}  also mass term $m_{H_{d}\widetilde{L},\eta}^{2}H_{d}^{\dag}\widetilde{L}+\mathrm{h.c.}$ mixing $H_{d}$ and slepton doublet $\widetilde{L}$ is generated by $\eta$ couplings.\footnote{After EWSB it may induce non-zero vev for sneutrino \cite{Barbier:2004ez}.} The explicit form of $m_{H_{d}\widetilde{L},\eta}^{2}$ can be found in the Appendix \ref{appmix}.

The role of $\eta$ will be further analysed in the Section \ref{pheno}, where we 
focus on the simplest examples of models restricted by additional global $U(1)_{q}$ symmetry. That symmetry is introduced in the next section  to meet phenomenological bounds.

\subsection{Operators generating proton decay and $\mu/B_{\mu}$ terms}\label{dangerous}

Having specified all possible superpotential couplings of messengers, it is important to know what are obstructions in getting realistic 
low-energy phenomenology, and how they are related to the discussed couplings \eqref{WY}. The obstacles one can face are 
e.g. rapid proton decay, absence of proper EWSB ($\mu/B_\mu$ problem \cite{Dvali:1996cu}), or R-parity violating soft terms in the Lagrangian \cite{primer}. In this Section we shall comment on such dangerous operators generated by \eqref{WY} at tree- and 1-loop level. 

When we discussed possible messenger couplings \eqref{WY}, gauge invariance and renormalizability were used as the only selection rules. Hence the tree-level dimension 4 operator $\phifb\phifb\phit|_{\t^2}$ could as well be present in the Lagrangian of the visible sector. However, it is well-known fact that this term would lead to rapid proton decay. So, first of all, it is necessary to ensure that it cannot appear in the superpotential. The  baryon/lepton number violation would also be induced by dimension 5 operators: $\phifb\phit^{3}/M_{GUT}|_{\t^2}$ or $\phifb\phit^{3}/M|_{\t^2}$. The latter appears after integrating out messengers. Its source 
in the model under consideration is the tree-level exchange of $(\Yf,\Yfb)$ messengers. It is clear that if matter-matter-messenger couplings $h_{2}$ and $h_{6}$ (cf. \eqref{WII}) occur simultaneously then they generate at the tree-level the following effective operator
\beqa\label{eff5}
\left.\frac{h_{2}h_{6}}{M}\phifb\phit\phit\phit\right|_{\t^2}.
\eeqa
To meet experimental bounds, at least one of these couplings must be highly suppressed such that $h_{2}h_{6}\lesssim10^{-26+t_M}$, where $t_{M}=\log_{10}(M/1\,\mathrm{GeV})$ \cite{Dudas:2010zb}. Integrating out messengers results also in corrections to the K\"ahler potential which may violate baryon/lepton number. In the discussed model, only the following operator of dimension 6 is relevant for the phenomenology:
\beq\label{six}
\left.\frac{h_{2}^{2}}{M^{2}}\phit^{\dag}\phit^{\dag}\phit\phit\right|_{\t^2\overline{\theta}^2}.
\eeq
If it is not suppressed then it lead to rapid proton decay, mainly in the channel $p\rightarrow\pi^{0}e^{+}$.  To satisfy the lower limit on the proton lifetime, $h_{2}$ has to fulfill
$h_{2}\lesssim10^{-16+t_{M}}$ \cite{Buchmuller:2004eg}.   

The second serious issue is the $\mu/B_{\mu}$ problem \cite{Dvali:1996cu}. We shall assume that there is no $\mu \Hf\Hfb$ term in the superpotential, and mass term for Higgs superfields is generated via the following correction to the K\"ahler potential
\beq\label{muX}
\frac{c_\mu}{M_{GUT}}X^\dag \Hf\Hfb
\eeq
when $F$-term of spurion superfield $X$ gets vev ($c_{\mu}$ is a coupling constant of order one).  
To avoid $\mu/B_{\mu}$ problem, we require that at the same time a term of the form $X^{\dag}X\Hf\Hfb/M_{GUT}^{2}$ is absent.  However, $\mu$ and $B_{\mu}$ are also generated 
at 1-loop by messenger-matter couplings i.e. the following corrections to the K\"ahler potential appear after integrating out messengers: 
\beqa\label{muBmueff}
\frac{c'_{\mu}}{M}X^{\dag}\Hf\Hfb\quad\textrm{and}\quad\frac{c'_{B_{\mu}}}{M^{2}}X^{\dag}X \Hf\Hfb,
\eeqa
where $c'_{\mu,B_{\mu}}\sim(h_{7}h_{9}+h_{11}h_{12})/(4\pi)^{2}\sim\mathcal{O}(10^{-2})$, which lead to $\mu^2\ll B_{\mu}$ when $F$-term of $X$ gets vev. 
Therefore, to get proper electroweak symmetry breaking with $\mu$ generated by \eqref{muX}, 
one has to suppress $c'_{\mu,B_{\mu}}$ \cite{Craig:2012xp}.

To ensure those phenomenological constraints without fine-tuning parameters, 
it is necessary to impose additional selection rules which restrict structure of the Lagrangian. 
It seems that the most handy and economical way to achieve it is to 
add extra global $U(1)_{q}$ symmetry. Such solution is realized  e.g. in F-theory GUT models \cite{Jelinski:2011xe,Heckman:2009mn,Pawelczyk:2010xh,Dolan:2011aq}, and in models which use  Froggatt-Nielsen mechanism to address Yukawas hierarchy problem. 

In the rest of the paper, we exploit that idea, and analyse what are necessary conditions to forbid dangerous operators which were discussed above. Then we shall examine what are the low-energy predictions of the simplified model in which  $U(1)_{q}$ symmetry dictates  the structure of the superpotential.  

\subsection{$U(1)_{q}$ symmetry}\label{U(1)}
As introduced in the previous section, to get rid of rapid proton decay and the $\mu/B_{\mu}$ problem we shall 
invoke extra global $U(1)_{q}$ symmetry and appropriately choose charges of the visible sector fields $\Phi$, messengers $(Y,\overline{Y})$ and the spurion $X$ . The requirements discussed in Section \ref{dangerous} can be rephrased as follows:
(a) in the superpotential of the visible sector there is no $\phifb\phifb\phit$ term, but the standard Yukawa couplings $\Hf\phit\phit$ and $\Hfb\phifb\phit$ are present, 
(b) supersymmetric mass term for the Higgses is forbidden, however they couple to $X^{\dag}$ in the K\"ahler potential. One of the ways to satisfy those conditions is to assign the following charges to the fields:
\beqa\label{charges}
q_{\Hfb}=k,\quad q_{\phifb}=l,\quad q_{\Hf}=2(k+l),\quad q_{\phit}=-(k+l),\quad q_X=3k+2l,
\eeqa
where $k$ and $l$ are nonequal integers such that charge of the $X$ field (i.e. $q_X$) is different from 0. 
Moreover, extra condition comes from spurion--messenger couplings \eqref{WX}. They   
are allowed only if the following relations hold:
\beq
q_{Y_{\overline{\mathsf{R}}}}=-(3k+2l)-q_{Y_\mathsf{R}}, 
\eeq
where $\mathsf{R}=5,10$ denotes representation of $SU(5)$.
Hence analysis of the possible $\Phi\Phi\sY$, $\Phi\sY\sY$ and $\sY\sY\sY$ couplings boils down to inspecting charges of $\Yf$ and $\Yt$. Let us remark that if they are set as in \eqref{charges} then operator $\phifb\phit^{3}/M_{GUT}|_{{\t}^2}$ is ruled out. Furthermore, condition (a) is basically equivalent to the requirement that $\phifb$ and $\Hfb$ have different $U(1)_{q}$ charges, what prevents them from mixing in the kinetic terms. It can also be checked that when these fields have different charges then \eqref{WY} does not lead to R-parity violating $A$-term (cf. Table \ref{tabAcoeff}). Moreover, it should be emphasized that operators \eqref{eff5} and \eqref{muBmueff} are also ruled out in spite of the fact that in the effective theory, below scale $M$,  global $U(1)_{q}$ symmetry is spontaneously broken by the vev of the lowest component of $X$. The reason is that they would arise from couplings $\phifb\phit^3/X$ and $\Hf\Hfb X^{\dag}/X$, respectively, which are forbidden in the parent theory provided (a) and (b) are satisfied. On the other hand, $U(1)_{q}$ symmetry protects the proton from decay via dimension 6 operator \eqref{six} only if 
\beq
q_{\Yf}\neq 2(k+l).
\eeq
Finally, mass term $\Hf\phifb$ in the superpotential which would cause appearance of $\phifb\phifb\phit$ operator after redefinition of matter fields should also be forbidden. To ensure this additional condition, we require that
\beq\label{2k3l}
2k+3l\neq0. 
\eeq

\section{Phenomenology of the simplest models with $\sY\sY\sY$}\label{pheno}

In this Section we study how low-energy predictions of the discussed extended GMSB models depend on messenger couplings \eqref{WYYY},  taking into account restrictions imposed by $U(1)_q$ symmetry. We shall use discussed above set of constraints (\ref{charges})-(\ref{2k3l}) to {\color{blue}} select the simplest models involving marginal interactions of messengers. Analysis of more complicated cases is straightforward, and will be given elsewhere. The cases with only $\eta$ couplings allowed give at the leading order the same results as the standard GMSB model with the effective number of messengers equal to 4, and will not be investigated here. It is also straightforward to check that there is no charge assignment which allows for model with only one $h$ and no $\eta$ coupling. 
On the other hand, taking into account (\ref{charges})-(\ref{2k3l}), it is easy to find out that there are only two possible scenarios which include one $h$ and one $\eta$. They are realized when superfields have $U(1)_{q}$ charges shown in the first (I) and the second (II) row of the table in the Appendix \ref{appU(1)q}. Then the allowed couplings are, respectively, 
\beq
\textrm{(I)}\quad(h_{8},\eta_4)\quad\quad\textrm{or}\quad\quad\textrm{(II)}\quad(h_{14},\eta_{2}).
\eeq
Surprisingly, there is a lot of ways to assign charges which allow for two $h$ and one $\eta$ interaction but not all of them are relevant for phenomenology. We shall examine two of them which lead to the biggest $A_t$-terms in that class of models (cf. Table \ref{tabAcoeff}). Namely, we analyse models in which $U(1)_{q}$ admits the following couplings:
\beq
\textrm{(III)}\quad(h_8,h_{11},\eta_2)\quad\quad\textrm{or}\quad\quad\textrm{(IV)}\quad(h_8,h_{11},\eta_4).
\eeq
Charges corresponding to these cases are displayed in the third (III) and fourth (IV) row of the table in the Appendix \ref{appU(1)q}.

\begin{figure}[!h]
\psfrag{st0}{{\footnotesize$\widetilde{\tau}_1$}}
\psfrag{se0}{{\footnotesize$\widetilde{e}_1$}}
\psfrag{st}{{\footnotesize$\widetilde{\tau}_1$}}
\psfrag{se}{{\footnotesize$\widetilde{e}_1$}}
\psfrag{bi}{{\footnotesize$\widetilde{\chi}_1^{0}$}}
\begin{center}
\includegraphics[scale=0.8]{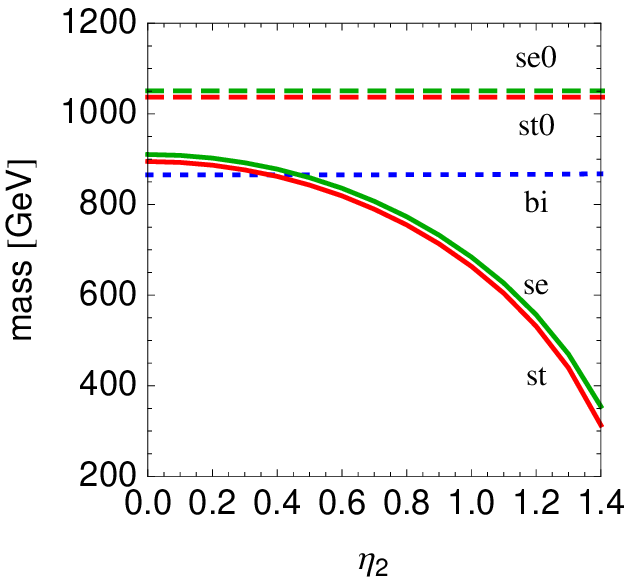}
\includegraphics[scale=0.8]{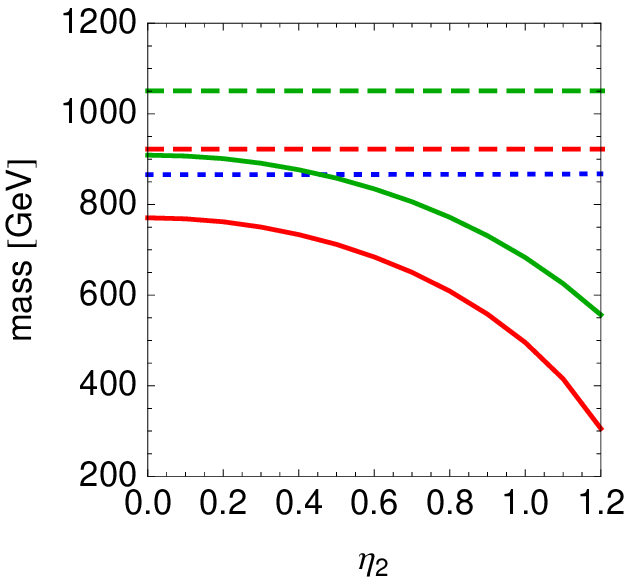}
\includegraphics[scale=0.8]{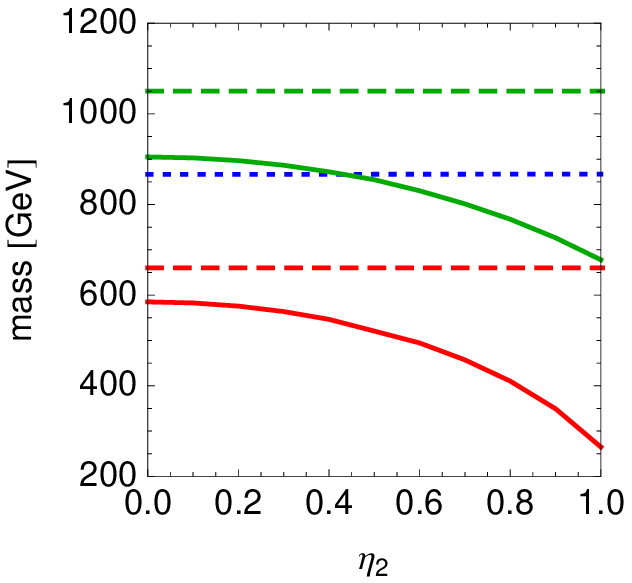}
\parbox{16.5cm}{
\caption{
\small Plot of the $\widetilde{\tau}_1$ (red, solid lines) and $\widetilde{e}_1$ (green, solid lines)  mass vs. $\eta_2$ coupling for $\tan\beta=10$ (left plot), $\tan\beta=30$ (middle plot) and $\tan\beta=50$ (right plot). Blue, dotted lines represent mass of lightest neutralino (bino). $h_{14}$ is set to 1.2, while $\xi$ scale is $1.6\times 10^5$ GeV. Dashed lines show masses of the particles when $h_{14}=\eta_2=0$, which corresponds to the standard GMSB case. The plots are symmetric under $\eta_{2}\rightarrow-\eta_{2}$ because of $m_{Q,\eta}^{2}\sim\eta_{2}^{2}$. 
In this case $\widetilde{\tau}_1$ and $\widetilde{e}_1$ are mostly right-handed.}
\label{fig:h14eta2}
}
\end{center}
\end{figure}

For the cases (I)-(IV) listed above,  
we adopt the initial conditions for the soft SUSY breaking terms presented in Sections \ref{softh} and \ref{softeta},
and compute the low-energy spectrum and the electroweak symmetry breaking
with an appropriately modified {\tt SuSpect} code \cite{Djouadi:2002ze}. 
Approximation of vanishing Yukawa couplings of the first two generations of fermions is used, so
the MSSM mass spectra we obtain are degenerate for these generations. In the following,
we shall call sfermions of the first and second generations with the name of the first generation (see e.g. Figure \ref{fig:h14eta2}).

A number of constraints is imposed on the obtained mass spectra. We require that the scalar potential
is bounded from below, and that there are no low lying color or charge breaking minima. All cases with tachyons in the spectrum, 
Higgs boson mass smaller than 123 GeV or bigger than 127 GeV 
and $\mathrm{BR}(b\to s\gamma)$ lying outside the 2$\sigma$ range
$(2.87-4.33)\times 10^{-4}$ are discarded during the analysis. 
We keep only models
in which the squark and gluino masses lie within the allowed $95\%$ CL range determined for a
simplified setup in Ref. \cite{Aad:2012fqa}.

In the case (I) the part of the superpotential which contains marginal couplings of the messengers is of the following form (cf. \eqref{WY}):
\beq
W_{\sY}=h_{8}\phit\Yf\Yt+\frac{1}{2}\eta_{4}\Yfb\Yfb\Yt.
\eeq
Although in such setup the left-right stop mixing would be large enough (cf. \eqref{Aterms} and Table \ref{tabAcoeff}) to get $m_{h^{0}}\sim125$ GeV even for $\xi\sim10^{5}$ GeV,  this case is not phenomenologically satisfactory as here supersymmetric mass term $M'\phifb\Yf$ is allowed. If $M'\sim M_{GUT}$, then the mass of the heavy combination of $\Yfb$ and $\phifb$ is of order GUT scale. As a result, soft terms generated by gauge mediation mechanism are of comparable size as those induced by gravity. Hence this case will not be analysed further. 

In the case (II)
only couplings $h_{14}$ and $\eta_2$ are allowed by $U(1)_{q}$ symmetry i.e.  
(cf. \eqref{WY}):
\beq
W_{\sY}=\frac{1}{2}h_{14}\phit\Yfb\Yfb+\frac{1}{2}\eta_{2}\Yfb\Ytb\Ytb.
\eeq
To get the lightest Higgs mass $m_{h^0}\approx125$ GeV in this simplified model, one has to set $\xi$ $\approx1.6\times 10^5$ GeV. The fact that  $m_{h^0}$ is not enhanced 
significantly by $h_{14}$ coupling 
can be traced back
to relatively small value of $C_{14}^{(t)}$ in the $A_t$-term \eqref{Aterms}.  The choice of $\xi$ results in rather large values of sparticles masses. Bino/wino/gluino masses are about $0.9/1.6/4.0$ TeV, and they hardly depend on $\eta_{2}$. On the other hand, masses of other neutralinos and charginos, Higgses (beside the lightest one), squarks and heavier sleptons do vary, but stay above 1.5 TeV when $\eta_{2}$ changes in the range $0-1.4$ and $h_{14}$ is fixed to $1.2$. However, it turns out that $\eta_2$ coupling influences NLSP/NNLSP pattern.  The lighter sleptons masses are sensitive to that messenger coupling, and  can be as low as 400 GeV for $h_{14},\eta_2\sim1$. Figure  \ref{fig:h14eta2} displays how masses of sleptons $\widetilde{\tau}_{1}$, $\widetilde{e}_{1}$ depend on the value of $\eta_2$. Increasing $\eta_2$ above $0.4-0.5$ changes NLSP/NNLSP pattern from $\widetilde{B}/\widetilde{\tau}_1$ to $\widetilde{\tau}_{1}/\widetilde{B}$ and then to $\widetilde{\tau}_{1}/\widetilde{e}_{1}$ (with nearly degenerated masses) for small $\tan \beta$, and from $\widetilde{\tau}_{1}/\widetilde{B}$ to $\widetilde{\tau}_{1}/\widetilde{e}_{1}$ for moderate and large values of $\tan\beta$. 
Such behaviour can be explained as follows. 

In this simplified model influence of marginal coupling of three messengers $\eta_2$ on mass spectrum is rather moderate because $\eta_2$ affects only soft mass of left squarks doublet $\widetilde{Q}$. 
One can check (cf. Appendix \ref{CQ}) that $\eta_{2}$ contribution to \eqref{m2soft} can be written as
\beq\label{m2Qeta}
m_{\widetilde{Q},\eta}^2=6\a_{h_{14}}\a_{\eta_{2}}\frac{\xi^2}{16\pi^2},
\eeq
which is always non-negative. For natural choice of coupling constants (i.e. $h,\eta\sim1$) it is of the same order as the following correction to \eqref{m2soft} induced by $h_{14}$ coupling
\beq\label{m2Qh}
m_{\widetilde{Q},h}^{2}=\a_{h_{14}}\left(6\a_{h_{14}}-\frac{7}{15}\a_{1}-3\a_{2}-6\a_{3}\right)\frac{\xi^2}{16\pi^2},
\eeq
and also of the same order as the  standard GMSB contribution
\beqa
m_{\widetilde{Q},g}^2=\left(\frac{2}{15}\a_1^2+6\a_2^2+\frac{32}{3}\a_3^2\right)\frac{\xi^2}{16\pi^2}.
\eeqa
Decreasing masses of lighter sleptons is the consequence of enlarging left squark doublet soft mass by \eqref{m2Qeta} which results in speeding up the running of $m^2_{\widetilde{L}}$ and $m^2_{\widetilde{E}}$ via $D$-term contribution to their RGE. 
Note that the increase in the splitting of the lighter sleptons masses when $\tan\beta$ becomes larger and larger is a well-known effect 
related to enhancing non-diagonal mass terms by $\tan\beta$ for the third family of sleptons. For the first and the second family such effect is suppressed by very small Yukawas.   

\begin{figure}[!h]
\psfrag{st0}{{\footnotesize$\widetilde{\tau}_1$}}
\psfrag{se0}{{\footnotesize$\widetilde{e}_1$}}
\psfrag{st}{{\footnotesize$\widetilde{\tau}_1,\widetilde{\nu}_{\tau}$}}
\psfrag{se}{{\footnotesize$\widetilde{e}_1$}}
\psfrag{bi}{{\footnotesize$\widetilde{\chi}_1^{0}$}}
\psfrag{snu0}{{\footnotesize$\widetilde{\nu}_{\tau}$}}
\begin{center}
\includegraphics[scale=0.8]{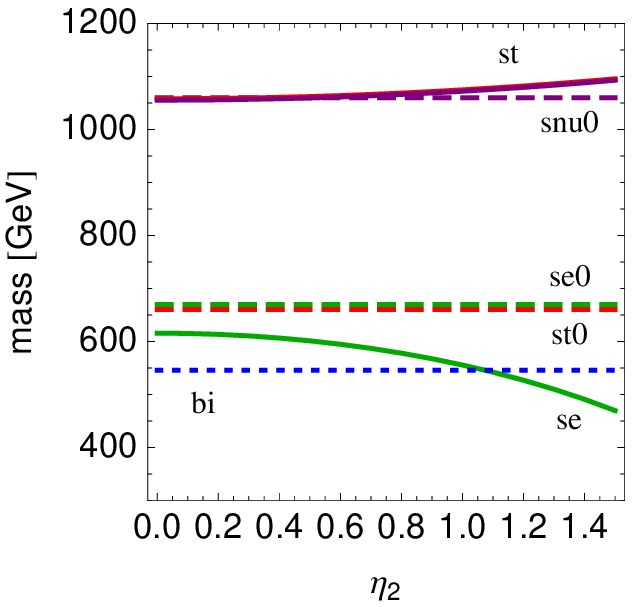}
\includegraphics[scale=0.8]{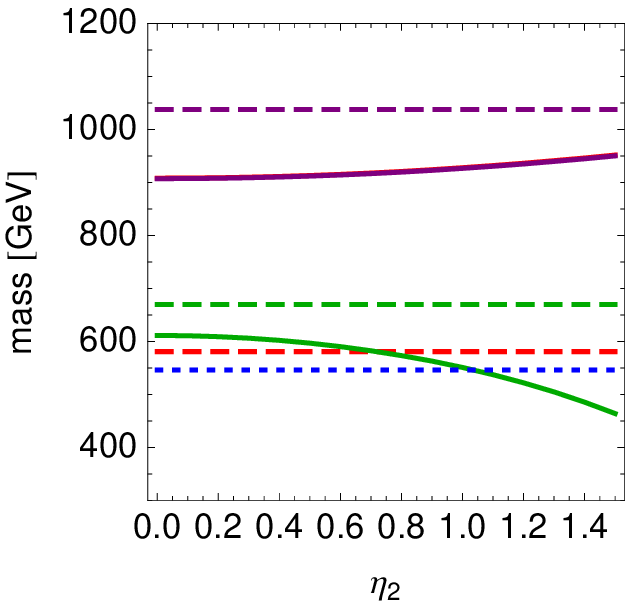}
\includegraphics[scale=0.8]{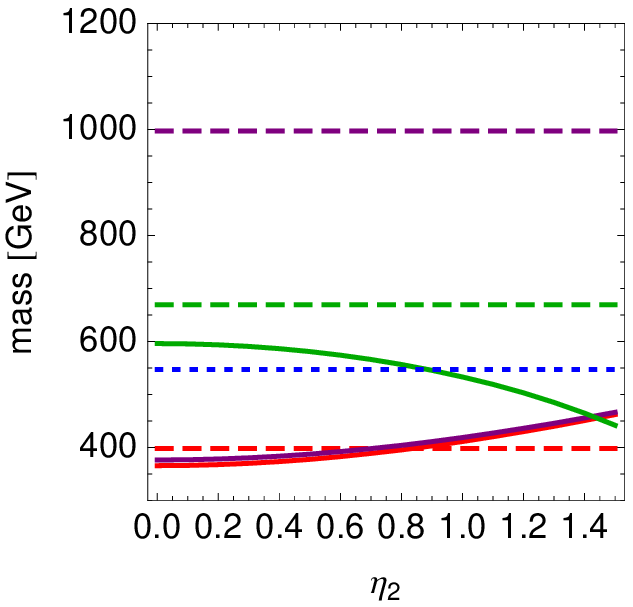}
\parbox{16.5cm}{
\caption{
\small Plot of the $\widetilde{\tau}_1$ (red, solid lines), $\widetilde{e}_1$ (green, solid lines) and $\widetilde{\nu}_{\tau}$ (purple, solid lines)  mass vs.  $\eta_2$ coupling for $\tan\beta=10$ (left plot), $\tan\beta=30$ (middle plot) and $\tan\beta=50$ (right plot). Blue, dotted lines represent mass of lightest neutralino (bino). The $\xi$ scale is chosen to be $10^5$ GeV. $h_8$ and $h_{11}$ are fixed  to 0.9 and 0.6, respectively, what results in $m_{h^0}\approx125$ GeV. Dashed lines show masses of the particles when $h_{8}=h_{11}=\eta_2=0$, which corresponds to the standard GMSB case. The plots are symmetric under $\eta_{2}\rightarrow-\eta_{2}$ because of $m_{H_u,\eta}^{2}\sim\eta_{2}^{2}$. Selectron and tau sneutrino masses are nearly degenerated. Here $\widetilde{e}_1$ is mostly right-handed, while $\widetilde{\tau}_1$ is mostly left-handed.}
\label{fig:h8h11eta2}
}
\end{center}
\end{figure}

In the framework defined by the third (III) choice of $U(1)_{q}$ charges, superpotential \eqref{WY} is of the following form
\beq
W_{\sY}=h_8\phit\Yf\Yt+h_{11}\Hf\Yf\Ytb+\frac{1}{2}\eta_2\Yfb\Ytb\Ytb, 
\eeq
and only corrections to $H_{u}$ mass are generated by messenger marginal couplings: 
\beq\label{m2Hueta}
m_{H_{u},\eta}^2=12\a_{h_{11}}\a_{\eta_2}\frac{\xi^2}{16\pi^2}.
\eeq
Here we choose $\xi=10^{5}$ GeV. Bino/wino/gluino masses are about $0.55/1.03/2.71$ TeV, and, as previously, they hardly depend on $\eta_{2}$. The first observation is that increasing $\eta_2$ gives smaller $\mu$ at the EWSB scale. For $\eta_2=0$ it is about $2.8$ TeV while for $\eta_2=1.2$ it drops to $2.4$ TeV. It is a consequence of extra contribution \eqref{m2Hueta} to $H_u$ soft mass. The aftereffect of changing $\mu$ is decreasing masses of heavy Higgses bosons from about 3.1 TeV to 2.6 TeV for small $\tan\beta$, and from about 2.4 TeV to 1.8 TeV for large $\tan\beta$. Similarly, heavier neutralinos and second chargino masses drop from $2.9$ TeV to $2.3$ TeV. Masses of the particles in the squark sector change only slightly. When $\eta_2$ grows they vary not more than a few percent and remain above 2 TeV. On the contrary, third family slepton masses are more sensitive to $\eta_2$ and $\tan\beta$ - see Figure \ref{fig:h8h11eta2}. The masses of the lighter stau and tau sneutrino are nearly degenerate, and grow when $\eta_2$ is increased while, at the same time, the heavier stau mass drops but it is always bigger than $2.3$ TeV. For small $\tan\beta$ the lighter stau mass increases from $1.06$ to $1.1$ TeV while for moderate and large $\tan\beta$ it raises from about $0.90$ to $0.95$ TeV  and from $0.36$ to $0.46$ TeV. The behaviour of the first and the second generation slepton masses is just the opposite. Namely, the mass of the lighter selectron reduces from about $0.6$ to $0.46$ TeV while mass of the heavier selectron and electron sneutrino increase from $1.07$ to $1.11$ TeV when $\eta_2$ changes from 0 to 1.4. 
Let us remark that here the lighter stau is mostly left-handed while the lighter selectron is mostly right-handed. 

As before, slepton masses are driven by $D$-term contribution to their RGE. \eqref{m2Qeta} increases these contributions for right sleptons and decreases for left sleptons. Since initial conditions at the messenger scale $M$  for the slepton masses are not altered by $\eta_{2}$, that results in decreasing masses of right sleptons and raising masses of left  sleptons. The reason why in this scenario left-handed stau is lighter than its right-handed counterpart is the contribution to slepton soft masses generated by $h_8$ and $h_{11}$:
\beq
m_{\widetilde{L},h}^2=-3\a_{h_8}\a_{y_{\tau}}\frac{\xi^2}{16\pi^2},\quad m_{\widetilde{E},h}^2=\left[-\left(\frac{28}{5}\a_1+16\a_3\right)\a_{h_8}+6\a_{h_{11}}\a_{h_8}+36\a_{h_8}^2\right]\frac{\xi^2}{16\pi^2}.
\eeq
Clearly, $m_{\widetilde{L},h}^2$ is amplified by large $\tan\beta$. It can be seen in the Figure  \ref{fig:h8h11eta2} that for large value of $\tan\beta$ and $\eta_{2}\lesssim1.4$ the NLSP is the lighter stau (which is mostly left-handed). The NNLSP is tau sneutrino and masses of both particles raise when $\eta_2$ increases. Simultaneously, the lighter selectron mass drops such that for $\eta_{2}\approx1.4$ it becomes close to the stau mass, and eventually NLSP/NNLSP pattern changes from $\widetilde{\tau}_1/\widetilde{\nu}_{\tau}$ to $\widetilde{e}_1/\widetilde{\tau}_1$.

\begin{figure}[!h]
\psfrag{st0}{{\footnotesize$\widetilde{\tau}_1$}}
\psfrag{se0}{{\footnotesize$\widetilde{e}_1,\widetilde{\tau}_1$}}
\psfrag{st}{{\footnotesize$\widetilde{\tau}_1,\widetilde{\nu}_{\tau}$}}
\psfrag{se}{{\footnotesize$\widetilde{e}_1$}}
\psfrag{bi}{{\footnotesize$\widetilde{\chi}_1^{0}$}}
\psfrag{snu0}{{\footnotesize$\widetilde{\nu}_{\tau}$}}
\begin{center}
\includegraphics[scale=0.8]{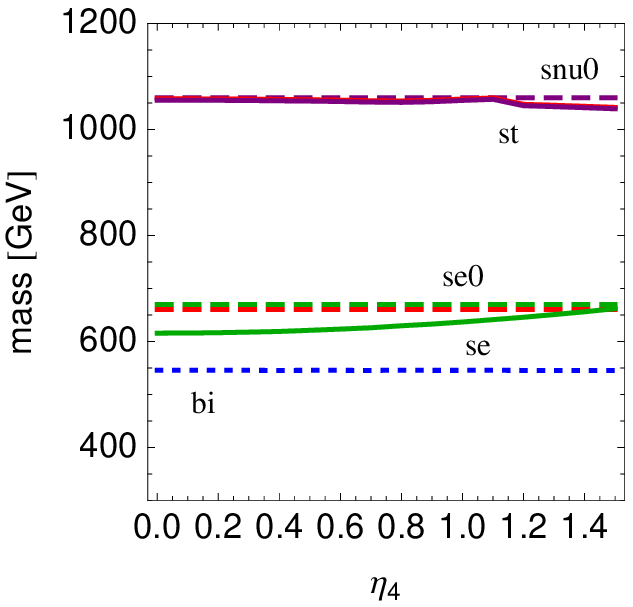}
\includegraphics[scale=0.8]{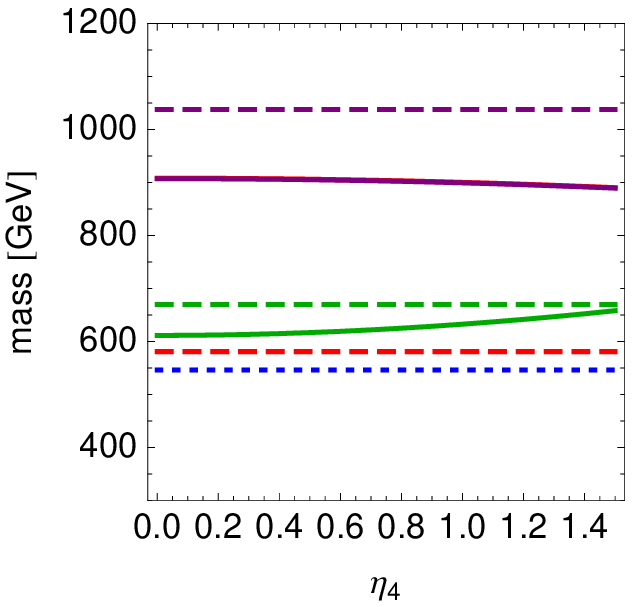}
\includegraphics[scale=0.8]{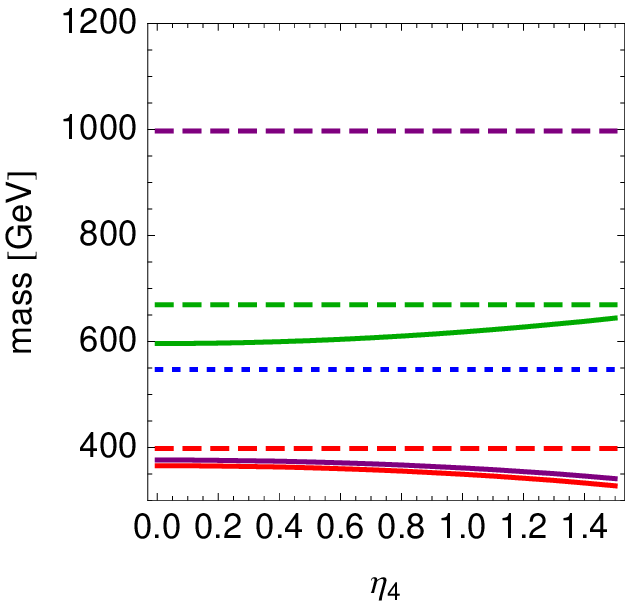}
\parbox{16.5cm}{
\caption{
\small Plot of the $\widetilde{\tau}_1$ (red, solid lines), $\widetilde{e}_1$ (green, solid lines) and $\widetilde{\nu}_{\tau}$ (purple, solid lines)  mass vs.  $\eta_4$ coupling for $\tan\beta=10$ (left plot), $\tan\beta=30$ (middle plot) and $\tan\beta=50$ (right plot). Blue, dotted lines represent mass of lightest neutralino (bino). The $\xi$ scale is chosen to be $10^5$ GeV. $h_8$ and $h_{11}$ are fixed  to 0.9 and 0.6, respectively, what results in $m_{h^0}\approx125$ GeV. Dashed lines show masses of the particles when $h_{8}=h_{11}=\eta_4=0$, which corresponds to the standard GMSB case. The plots are symmetric under $\eta_{4}\rightarrow-\eta_{4}$ because of $m_{Q,\eta}^{2}=m_{U,\eta}^{2}\sim\eta_{4}^{2}$. Selectron and tau sneutrino masses are nearly degenerated. Here $\widetilde{e}_1$ is mostly right-handed, while $\widetilde{\tau}_1$ is mostly left-handed.}
\label{fig:h8h11eta4}
}
\end{center}
\end{figure}

In the case (IV) superpotential is of the form
\beq
W_{\sY\sY\sY}=h_8\phit\Yf\Yt+h_{11}\Hf\Yf\Ytb+\frac{1}{2}\eta_4\Yfb\Yfb\Yt,
\eeq
and coupling of three messengers induces the following contributions to soft masses of left squarks and right up squark
\beq\label{m2Qeta4}
m_{\widetilde{Q},\eta}^2=m_{\widetilde{U},\eta}^2=2\a_{h_8}\a_{\eta_4}.
\eeq
Their influence on the mass spectrum can be described in the following way. Bino/wino/gluino masses are about $0.55/1.03/2.71$ TeV, and, as before, they hardly depend on $\eta_{4}$. On the other hand, behaviour of $\mu$ is different than in previous case. In this scenario $\mu$ increases  when $\eta_4$ raises. The same effect can be seen in the Higgs sector (beside the lightest Higgs scalar) and for heavier chargino and heavier neutralinos. All these masses become larger of about $150-200$ GeV and approach level of $3$ TeV. It is caused by $\eta_4$ contribution to soft masses of left squarks and right up squark. They enter RGE for the soft mass of $H_{u}$ what results in bigger value of $|m_{H_u}^2|$ at EWSB scale. 
Masses of the stops and lighter sbottom are the most sensitive to $\eta_4$. They increase of about $200$ GeV when $\eta_4$ is changed, and reach level of $2.5-3$ TeV. The first and second generation squarks masses hardly depend on $\eta_4$ value. On the other hand, the mass of the lighter stau and tau sneutrino drop when $\eta_4$ is increased while heavier stau mass raises at the same time - see Figure \ref{fig:h8h11eta4}. Tau sneutrino behaves similarly, and its mass is nearly degenerate with the lighter stau mass. Again, behaviour of the first and second generation slepton masses is just the opposite. Here the lighter stau is mostly left-handed while the lighter selectron is mostly left-handed. 

Analogously to the previous cases, additional contributions to squarks soft masses \eqref{m2Qeta4} decrease $D$-term contribution to RGE for right sleptons, and increase for left sleptons. That results in raising masses of right sleptons and decreasing masses of left sleptons at the EWSB scale. Left-handed stau is lighter than its right-handed counterpart from the same reason as in the case (III).

\section{Conclusions}

In this work we have studied extended GMSB model in which messenger sector consists of fields in fundamental and antisymmetric representation of $SU(5)$ (and their conjugates). 
We have shown that in such scenario superpotential couplings of three messengers \eqref{WYYY} induce additional contributions to the standard soft masses of scalars when they coexist with appropriate couplings between messengers and MSSM matter i.e. \eqref{WII} or \eqref{WI}. 
Namely, they generate 2-loop corrections to the soft masses when one of the messenger fields enter both $\sY\sY\sY$ and $\sY\Phi\Phi$ or $\sY\sY\Phi$ vertices. At the same time, they 
lead to neither additional $A$-terms 
nor 1-loop contributions to soft masses, what may be of some importance for low-scale SUSY breaking models. 
We have derived all 2-loop soft masses and 1-loop $A$-terms for the most general, marginal superpotential couplings \eqref{WY} allowed by gauge symmetry in the discussed model. 

It turns out that to fulfill phenomenological constraints it is necessary to impose extra selection rules on \eqref{WY}. Otherwise rapid proton decay or $\mu/B_\mu$ problem can occur. We deal with those issues by invoking additional global $U(1)_q$ symmetry.
The charge assignments which lead to the smallest number of allowed interaction terms were found.  Using derived corrections to the soft masses \eqref{m2eta}, we have performed analysis of the phenomenology of the models involving the smallest number of marginal couplings of three messengers. The main conclusion is that in those scenarios the lightest slepton masses are the most sensitive to $\eta$ couplings, which alter them not directly but only via $D$-term contribution to RGE running. We have shown that due to $\eta$, even for small $\tan\beta$ (see Figure \ref{fig:h14eta2} and Figure \ref{fig:h8h11eta2}, left plots), a stau or selectron can be lighter than the lightest neutralino, and have masses as low as $300-400$ GeV, which is close to recent LHC exclusion limit. Such situation is not typical for the GMSB model.

It would be worthwhile to extend analysis of the parameter space of the presented model, especially to the low $M$ region, and investigate whether marginal couplings of three messengers are relevant for realizing radiative EWSB.   
Moreover, one can check if there is any common NLSP/NNLSP pattern which emerge when one considers models with more $h$ and $\eta$ couplings allowed. Finally, while we assumed that messenger-matter couplings are hierarchical, it would be interesting to investigate full flavour structure of those interactions in the discussed model.

\subsubsection*{Acknowledgments}
\noindent \normalsize The author would like to thank 
J. Gluza, J. Pawe{\l}czyk and K. Turzy\'nski for useful discussions and comments. This work
was supported by the National Science Center under post-doctoral grant DEC-2012/04/S/ST2/00003 and in part by the Research Executive Agency (REA) of the European Union under the Grant Agreement number PITN-GA-2010-264564 (LHCPhenoNet).

\appendix

\section{$U(1)_{q}$ charges}
\label{appU(1)q}

Here we show $U(1)_{q}$ charge assignments which lead to the smallest number of allowed marginal couplings of messengers. In the first case (I)  only couplings $h_{8}$ and $\eta_{4}$ occur while the second choice of charges (II) results in the presence of $h_{14}$ and $\eta_{2}$. 
Cases (III) and (IV) correspond to models with two $h_A$ couplings and one $\eta_i$ coupling which accommodate the largest $A_t$-terms in that class of models. 
In the table below, the charges of the fields  are 
written as multiplicities of the smallest charge (denoted by $q_{1}$, $q_{2}$, $q_{3}$ and $q_{4}$ respectively).

{\footnotesize
\begin{center}
\begin{tabular}{ccc|c|c|c|c|c|c|c|c|}
\cline{3-11}
\multicolumn{2}{c|}{}& $\Hf$ & $\Hfb$ & $\phifb$ & $\phit$ & $\Yf$ & $\Yfb$ & $\Yt$ & $\Ytb$ & $X$\\
\cline{3-11}
\hline
\multicolumn{1}{|l}{(I)} & \multicolumn{1}{l|}{$(h_{8},\eta_{4})$} & $2q_{1}$ & $0$ & $q_{1}$ & $-q_{1}$ & $-q_{1}$ & $-q_{1}$ & $2q_{1}$ & $-4q_{1}$ & $2q_{1}$\\
\multicolumn{1}{|l}{(II)} & \multicolumn{1}{l|}{$(h_{14},\eta_{2})$} & $-8q_{2}$ & $-7q_{2}$  & $3q_{2}$ & $4q_{2}$ & $17q_{2}$ & $-2q_{2}$ & $14q_{2}$ & $q_{2}$ & $-15q_{2}$\\
\multicolumn{1}{|l}{(III)} & \multicolumn{1}{l|}{$(h_{8},h_{11},\eta_{2})$} & $2q_3$ & $-3q_3$ & $4q_3$ & $-q_3$ & $-q_3$ & $2q_3$ & $2q_3$ & $-q_3$ & $-q_3$\\
\multicolumn{1}{|l}{(IV)} & \multicolumn{1}{l|}{$(h_{8},h_{11},\eta_{4})$} & $14q_4$ & $-9q_4$ & $16q_4$ & $-7q_4$ & $-q_4$ & $-4q_4$ & $8q_4$ & $-13q_4$ & $5q_4$\\
\hline
\end{tabular}
\end{center}
}

\section{Numerical coefficients  in 2-loop soft masses}
\label{appCs}
In this Appendix we tabulate 
numerical values of the coefficients $C_{i,A}^{(\Phi)}$, $C_{i,(A,B,C)}^{(\Phi)}$ and $C_{i,(A,B,f)}^{(\Phi)}$ which appear in 2-loop corrections \eqref{m2eta} to sfermions soft masses induced by marginal couplings of three messengers \eqref{WYYY}. These coefficients are displayed in the tables below. Their rows are indexed by $i=1,2,3,4$.
On the other hand, columns are indexed either by $A$ or by triples: $(A,B,C)$ or $(A,B,f)$ depending on the case. The $C^{(\Phi)}_{i,A}$ coefficient can be found at the $i$-th row and $A$-th column. Likewise, $C^{(\Phi)}_{i,(A,B,C)}$ or $C^{(\Phi)}_{i,(A,B,f)}$ coefficients are located at the $i$-th row and column denoted by $(A,B,C)$ or $(A,B,f)$. The ranges of indices are: $A,B=1,\ldots,14$ and $f=t,b,\tau$. All $C^{(\Phi)}$'s which are not  listed here are zero.

\subsection{Higgs $H_{u}$}
\label{appHu}

{\footnotesize
\begin{center}
\begin{tabular}{cc|c|c|c|c|c|c|c|}
\cline{3-9}
&&\multicolumn{3}{c|}{$A$}&\multicolumn{2}{c|}{$(A,B,C)$}&\multicolumn{2}{c|}{$(A,B,f)$}\\
\cline{3-9}
&& $1$ & $7$ & $11$ & $(1,7,8)$ & $(1,7,14)$ & $(1,8,t)$ & $(1,14,t)$\\
\cline{3-9}
\cline{1-2}\cline{3-9}
\multicolumn{1}{|c|}{\multirow{4}{*}{$i$}}&\multicolumn{1}{c|}{$1$} & $18$ & $18$ & $12$ & $36$ & $0$ & $36$ & $0$\\
\multicolumn{1}{|c|}{}&\multicolumn{1}{c|}{$2$}                               & $0$   & $0$   & $12$ & $0$   & $0$ & $0$   & $0$\\
\multicolumn{1}{|c|}{}&\multicolumn{1}{c|}{$3$}                               & $0$   & $0$   & $12$ & $0$   & $0$ & $0$   & $0$\\
\multicolumn{1}{|c|}{}&\multicolumn{1}{c|}{$4$}                               & $3$   & $3$   & $0$   & $0$   & $6$ & $0$   & $6$\\
\hline
\end{tabular}
\end{center}
}

\subsection{Higgs $H_{d}$}

{\footnotesize
\begin{center}
\begin{tabular}{cc|c|c|c|c|c|c|c|c|}
\cline{3-10}
&&\multicolumn{4}{c|}{$A$}&\multicolumn{4}{c|}{$(A,B,C)$}\\
\cline{3-10}
& & $3$ & $4$ & $9$ & $12$ & $(4,8,12)$ & $(3,10,12)$ & $(3,12,13)$ & $(4,12,14)$\\
\hline
\multicolumn{1}{|c|}{\multirow{4}{*}{$i$}}&\multicolumn{1}{c|}{$1$} & $12$ & $0$   &  $0$  & $12$ & $24$ & $0$   & $0$   & $0$\\
\multicolumn{1}{|c|}{}&\multicolumn{1}{c|}{$2$}                               & $0$   & $12$ & $18$ & $12$ & $0$   & $24$ & $0$   & $0$\\
\multicolumn{1}{|c|}{}&\multicolumn{1}{c|}{$3$}                               & $0$   & $0$   & $3$   & $0$   & $0$   & $0$   & $0$   & $0$\\
\multicolumn{1}{|c|}{}&\multicolumn{1}{c|}{$4$}                               & $3$   & $9$   & $0$   & $12$ & $0$   & $0$   & $18$ & $6$\\
\hline
\end{tabular}
\end{center}
}
\vspace{2.5mm}
{\footnotesize
\begin{center}
\begin{tabular}{cc|c|c|c|c|c|c|c|c|}
\cline{3-9}
&&\multicolumn{7}{c|}{$(A,B,f)$}\\
\cline{3-9}
&& $(3,8,b)$ & $(3,8,\tau)$ & $(4,10,b)$ & $(4,10,\tau)$ & $(3,14,b)$ & $(4,13,b)$ & $(4,13,\tau)$\\
\hline
\multicolumn{1}{|c|}{\multirow{4}{*}{$i$}}&$1$ & $18$ & $6$   & $0$   & $0$ & $0$ & $0$   & $0$\\
\multicolumn{1}{|c|}{}&$2$                               & $0$   & $0$   & $18$ & $6$ & $0$ & $0$   & $0$\\
\multicolumn{1}{|c|}{}&$3$                               & $0$   & $0$   & $3$   & $0$ & $0$ & $0$   & $0$\\
\multicolumn{1}{|c|}{}&$4$                               & $0$   & $0$   & $0$   & $0$ & $6$ & $12$ & $6$\\
\hline
\end{tabular}
\end{center}
}

\subsection{Slepton doublet $\widetilde{L}$}

{\footnotesize
\begin{center}
\begin{tabular}{cc|c|c|c|c|c|c|c|c|c|c|c|c|}
\cline{3-13}
&&\multicolumn{5}{c|}{$A$}&\multicolumn{6}{c|}{$(A,B,C)$}\\
\cline{3-13}
& & $3$ & $5$ & $6$ & $10$ & $13$& $(6,8,13)$ & $(5,10,13)$ & $(3,9,13)$ & $(3,12,13)$ & $(5,13,13)$ & $(6,13,14)$\\
\hline
\multicolumn{1}{|c|}{\multirow{4}{*}{$i$}}&$1$ & $3$ & $9$ & $0$   & $0$   & $12$ & $24$ & $0$   & $0$ & $0$ & $0$   & $0$\\
\multicolumn{1}{|c|}{}&$2$                               & $0$ & $0$ & $12$ & $18$ & $12$ & $0$   & $18$ & $6$ & $0$ & $0$   & $0$\\
\multicolumn{1}{|c|}{}&$3$                               & $0$ & $0$ & $0$   & $3$   & $0$   & $0$   & $0$   & $0$ & $0$ & $0$   & $0$\\
\multicolumn{1}{|c|}{}&$4$                               & $0$ & $3$ & $9$   & $0$   & $12$ & $0$   & $0$   & $0$ & $6$ & $12$ & $6$\\
\hline
\end{tabular}
\end{center}
}

{
\footnotesize
\begin{center}
\begin{tabular}{cc|c|c|c|}
\cline{3-5}
&&\multicolumn{3}{c|}{$(A,B,f)$}\\
\cline{3-5}
& & $(3,8,\tau)$ & $(6,9,\tau)$ & $(6,12,\tau)$\\
\hline
\multicolumn{1}{|c|}{\multirow{4}{*}{$i$}}&\multicolumn{1}{c|}{$1$} & $6$ & $0$ & $0$\\
\multicolumn{1}{|c|}{}&\multicolumn{1}{c|}{$2$}                               & $0$ & $6$ & $0$\\
\multicolumn{1}{|c|}{}&\multicolumn{1}{c|}{$3$}                               & $0$ & $0$ & $0$\\
\multicolumn{1}{|c|}{}&\multicolumn{1}{c|}{$4$}                               & $0$ & $0$ & $6$\\
\hline
\end{tabular}
\end{center}
}

\subsection{Right stau $\widetilde{E}$}

{\footnotesize
\begin{center}
\begin{tabular}{cc|c|c|c|c|c|c|c|c|c|}
\cline{3-11}
&&\multicolumn{4}{c|}{$A$}&\multicolumn{1}{c|}{$(A,B,C)$}&\multicolumn{4}{c|}{$(A,B,f)$}\\
\cline{3-11}
& & $2$ & $4$ & $6$ & $8$ & $(2,8,8)$ & $(4,10,\tau)$ & $(6,9,\tau)$ & $(4,13,\tau)$ & $(6,12,\tau)$\\
\hline
\multicolumn{1}{|c|}{\multirow{4}{*}{$i$}}&\multicolumn{1}{c|}{$1$} & $9$ & $0$ & $0$ & $18$ & $18$ & $0$   & $0$   & $0$   & $0$\\
\multicolumn{1}{|c|}{}&\multicolumn{1}{c|}{$2$}                               & $0$ & $6$ & $6$ & $0$   & $0$   & $12$ & $12$ & $0$   & $0$\\
\multicolumn{1}{|c|}{}&\multicolumn{1}{c|}{$3$}                               & $6$ & $0$ & $0$ & $6$   & $0$   & $0$   & $0$   & $0$   & $0$\\
\multicolumn{1}{|c|}{}&\multicolumn{1}{c|}{$4$}                               & $0$ & $6$ & $6$ & $0$   & $0$   & $0$   & $0$   & $12$ & $12$\\
\hline
\end{tabular}
\end{center}
}

\subsection{Squark doublet $\widetilde{Q}$}
\label{CQ}
{\footnotesize
\begin{center}
\begin{tabular}{cc|c|c|c|c|c|c|c|c|c|c|c|c|c|c|}
\cline{3-16}
&&\multicolumn{6}{c|}{$A$}&\multicolumn{8}{c|}{$(A,B,C)$}\\
\cline{3-16}
& & $1$ & $2$ & $4$ & $6$ & $8$ & $14$ & $(1,7,8)$ & $(2,8,8)$ & $(6,10,14)$ & $(4,9,14)$ & $(1,8,11)$ & $(4,12,14)$ & $(6,13,14)$ & $(2,8,14)$\\
\hline
\multicolumn{1}{|c|}{\multirow{4}{*}{$i$}}&\multicolumn{1}{c|}{$1$} & $3$ & $9$ & $0$ & $0$ & $18$ & $0$ & $6$  & $18$ & $0$   & $0$ & $0$ & $0$ & $0$ & $0$\\
\multicolumn{1}{|c|}{}&\multicolumn{1}{c|}{$2$}                               & $0$ & $0$ & $3$ & $6$ & $0$   & $6$ & $0$  & $0$   & $12$ & $6$ & $0$ & $0$ & $0$ & $0$\\
\multicolumn{1}{|c|}{}&\multicolumn{1}{c|}{$3$}                               & $0$ & $7$ & $0$ & $0$ & $7$   & $0$ & $0$  & $0$   & $0$   & $0$ & $6$ & $0$ & $0$ & $0$\\
\multicolumn{1}{|c|}{}&\multicolumn{1}{c|}{$4$}                               & $0$ & $6$ & $2$ & $5$ & $2$   & $5$ & $0$  & $0$   & $0$   & $0$ & $0$ & $6$ & $10$ & $4$\\
\hline
\end{tabular}
\end{center}
}
\vspace{2mm}
{
\footnotesize
\begin{center}
\begin{tabular}{cc|c|c|c|c|c|c|c|}
\cline{3-9}
&&\multicolumn{7}{c|}{$(A,B,f)$}\\
\cline{3-9}
& & $(2,7,t)$ & $(1,8,t)$ & $(4,10,b)$ & $(6,9,b)$ & $(2,11,t)$ & $(4,13,b)$ & $(6,12,b)$\\
\hline
\multicolumn{1}{|c|}{\multirow{4}{*}{$i$}}&\multicolumn{1}{c|}{$1$} &  $6$  &  $6$  &  $0$ & $0$ & $0$ & $0$ & $0$ \\
\multicolumn{1}{|c|}{}&\multicolumn{1}{c|}{$2$}                               &  $0$  &  $0$  &  $6$ & $6$ & $0$ & $0$ & $0$ \\
\multicolumn{1}{|c|}{}&\multicolumn{1}{c|}{$3$}                               &  $0$  &  $0$  &  $0$ & $0$ & $6$ & $0$ & $0$ \\
\multicolumn{1}{|c|}{}&\multicolumn{1}{c|}{$4$}                               &  $0$  &  $0$  &  $0$ & $0$ & $0$ & $4$ & $6$ \\
\hline
\end{tabular}
\end{center}
}

\subsection{Right stop $\widetilde{U}$}

{\footnotesize
\begin{center}
\begin{tabular}{cc|c|c|c|c|c|c|c|c|c|c|c|c|}
\cline{3-14}
&&\multicolumn{4}{c|}{$A$}&\multicolumn{4}{c|}{$(A,B,C)$}&\multicolumn{4}{c|}{$(A,B,f)$}\\
\cline{3-14}
& & $1$ & $2$ & $6$ & $8$ & $(1,7,8)$ & $(2,8,8)$ & $(1,8,11)$ & $(2,8,14)$ & $(2,7,t)$ & $(1,8,t)$ & $(2,11,t)$ & $(1,14,t)$\\
\hline
\multicolumn{1}{|c|}{\multirow{4}{*}{$i$}}&\multicolumn{1}{c|}{$1$} & $6$ & $9$ & $0$ & $18$ & $12$ & $18$ & $0$   & $0$ & $12$ & $12$ & $0$   & $0$\\
\multicolumn{1}{|c|}{}&\multicolumn{1}{c|}{$2$}                               & $0$ & $0$ & $6$ & $0$   & $0$   & $0$   & $0$   & $0$ & $0$   & $0$   & $0$   & $0$\\
\multicolumn{1}{|c|}{}&\multicolumn{1}{c|}{$3$}                               & $0$ & $8$ & $0$ & $8$   & $0$   & $0$   & $12$ & $0$ & $0$   & $0$   & $12$ & $0$\\
\multicolumn{1}{|c|}{}&\multicolumn{1}{c|}{$4$}                               & $2$ & $0$ & $4$ & $2$   & $0$   & $0$   & $0$   & $4$ & $0$   & $0$   & $0$   & $4$\\
\hline
\end{tabular}
\end{center}
}

\subsection{Right sbottom $\widetilde{D}$}
\label{appD}

{\footnotesize
\begin{center}
\begin{tabular}{cc|c|c|c|c|c|c|c|c|c|c|c|c|}
\cline{3-13}
&&\multicolumn{5}{c|}{$A$}&\multicolumn{6}{c|}{$(A,B,C)$}\\
\cline{3-13}
& & $3$ & $5$ & $6$ & $10$ & $13$& $(6,8,13)$ & $(5,10,13)$ & $(3,9,13)$ & $(3,12,13)$ & $(5,13,13)$ & $(6,13,14)$\\
\hline
\multicolumn{1}{|c|}{\multirow{4}{*}{$i$}}&$1$ & $6$ & $6$ & $0$   & $0$   & $12$ & $24$ & $0$   & $0$   & $0$   & $0$   & $0$\\
\multicolumn{1}{|c|}{}&$2$                               & $0$ & $0$ & $12$ & $18$ & $12$ & $0$   & $12$ & $12$ & $0$   & $0$   & $0$\\
\multicolumn{1}{|c|}{}&$3$                               & $0$ & $0$ & $0$   & $4$   & $0$   & $0$   & $0$   & $0$   & $0$   & $0$   & $0$\\
\multicolumn{1}{|c|}{}&$4$                               & $2$ & $2$ & $10$ & $0$   & $12$ & $0$   & $0$   & $0$   & $12$ & $12$ & $4$\\
\hline
\end{tabular}
\end{center}
}

{\footnotesize
\begin{center}
\begin{tabular}{cc|c|c|c|c|}
\cline{3-6}
&&\multicolumn{4}{c|}{$(A,B,f)$}\\
\cline{3-6}
& & $(3,8,b)$ & $(6,9,b)$ & $(3,14,b)$ & $(6,12,b)$\\
\hline
\multicolumn{1}{|c|}{\multirow{4}{*}{$i$}}&\multicolumn{1}{c|}{$1$} & $12$ & $0$   & $0$ & $0$\\
\multicolumn{1}{|c|}{}&\multicolumn{1}{c|}{$2$}                               & $0$   & $12$ & $0$ & $0$\\
\multicolumn{1}{|c|}{}&\multicolumn{1}{c|}{$3$}                               & $0$   & $0$   & $0$ & $0$\\
\multicolumn{1}{|c|}{}&\multicolumn{1}{c|}{$4$}                               & $0$   & $0$   & $4$ & $12$\\
\hline
\end{tabular}
\end{center}
}

\subsection{$H_d-\widetilde{L}$ mixing}\label{appmix}
For the completeness of the discussion, here we show numerical coefficients which appear in the contributions to 2-loop mixing masses of sleptons and down Higgses: $m_{H_{d}\widetilde{L}}^{2}H_{d}^{\dag}\widetilde{L}+\mathrm{h.c.}$ generated by marginal couplings of the messengers \eqref{WYYY}. $m_{H_{d}\widetilde{L}}^{2}$ can be written as follows: 
\beqa\label{m2mix}
m^{2}_{H_{d}\widetilde{L},\eta}&=&\frac{\xi^2}{16\pi^2}\sum_{iA\leq B\leq Cf}\left(C^{(H_{d}\widetilde{L})}_{i,(A,B)}(\a_{\eta_i}^{2}\a_{h_A}\a_{h_{B}})^{1/2}+C^{(H_{d}\widetilde{L})}_{i,(A,B,C)}(\a_{\eta_{i}}\a_{h_A}\a_{h_B}\a_{h_C})^{1/2}\right.\nn\\
&&\phantom{\frac{\xi^2}{16\pi^2}\sum_{iA\leq B\leq Cf}}\left.+C^{(H_{d}\widetilde{L})}_{i,(A,B,f)}(\a_{\eta_{i}}\a_{h_A}\a_{h_B}\a_{f})^{1/2}\right).
\eeqa
Coefficients $C^{(H_{d}\widetilde{L})}_{i,(A,B)}$, $C^{(H_{d}\widetilde{L})}_{i,(A,B,C)}$ and $C^{(H_{d}\widetilde{L})}_{i,(A,B,f)}$ which appear in \eqref{m2mix} are displayed in the tables below. Analogously to \ref{appHu}--\ref{appD}, the $C^{(H_{d}\widetilde{L})}_{i,(A,B)}$ coefficient can be found at the $i$-th row and $(A,B)$-th column of the appropriate table. Likewise, $C^{(H_{d}\widetilde{L})}_{i,(A,B,C)}$ or $C^{(H_{d}\widetilde{L})}_{i,(A,B,f)}$ coefficients are located at the $i$-th row and column denoted by $(A,B,C)$ or $(A,B,f)$. The ranges of indices are as in the previous cases. All $C^{(H_{d}\widetilde{L})}$'s which are not  listed here are zero.

{\footnotesize
\begin{center}
\begin{tabular}{cc|c|c|c|c|}
\cline{3-6}
&&\multicolumn{4}{c|}{$(A,B)$}\\
\cline{3-6}
& & $(12,13)$ & $(3,5)$ & $(4,6)$ & $(9,10)$\\
\hline
\multicolumn{1}{|c|}{\multirow{4}{*}{$i$}}&$1$ & $12$ & $9$ & $0$   & $0$\\
\multicolumn{1}{|c|}{}&$2$                               & $12$ & $0$ & $12$ & $18$\\
\multicolumn{1}{|c|}{}&$3$                               & $0$   & $0$ & $0$   & $3$\\
\multicolumn{1}{|c|}{}&$4$                               & $12$ & $3$ & $9$   & $0$\\
\hline
\end{tabular}
\end{center}
}

{\footnotesize
\begin{center}
\begin{tabular}{cc|c|c|c|c|c|c|c|c|c|c|c|}
\cline{3-12}
&&\multicolumn{10}{c|}{$(A,B,C)$}\\
\cline{3-12}
&& $(4,8,13)$ & $(6,8,12)$ & $(3,10,13)$ & $(5,10,12)$ & $(3,9,12)$ & $(3,12,12)$ & $(3,13,13)$ & $(4,13,14)$ & (5,12,13) & (6,12,14)\\
\hline
\multicolumn{1}{|c|}{\multirow{4}{*}{$i$}}&$1$ & $12$ & $12$ & $0$   & $0$ & $0$ & $0$ & $0$ & $0$ & $0$ & $0$\\
\multicolumn{1}{|c|}{}&$2$                               & $0$   & $0$   & $12$ & $9$ & $3$ & $0$ & $0$ & $0$ & $0$ & $0$\\
\multicolumn{1}{|c|}{}&$3$                               & $0$   & $0$   & $0$   & $0$ & $0$ & $0$ & $0$ & $0$ & $0$ & $0$\\
\multicolumn{1}{|c|}{}&$4$                               & $0$   & $0$   & $0$   & $0$ & $0$ & $3$ & $9$ & $3$ & $6$ & $3$\\
\hline
\end{tabular}
\end{center}
}

{\footnotesize
\begin{center}
\begin{tabular}{cc|c|c|c|c|c|c|c|c|c|}
\cline{3-10}
&&\multicolumn{8}{c|}{$(A,B,f)$}\\
\cline{3-10}
&& $(5,8,b)$ & $(6,10,b)$ & $(6,10,\tau)$ & $(4,9,\tau)$ & $(5,14,b)$ & $(6,13,b)$ & $(4,12,\tau)$ & $(6,13,\tau)$\\
\hline
\multicolumn{1}{|c|}{\multirow{4}{*}{$i$}}&$1$ & $9$ & $0$ & $0$ & $0$ & $0$ & $0$ & $0$ & $0$\\
\multicolumn{1}{|c|}{}&$2$                               & $0$ & $9$ & $3$ & $3$ & $0$ & $0$ & $0$ & $0$\\
\multicolumn{1}{|c|}{}&$3$                               & $0$ & $0$ & $0$ & $0$ & $0$ & $0$ & $0$ & $0$\\
\multicolumn{1}{|c|}{}&$4$                               & $0$ & $0$ & $0$ & $0$ & $3$ & $6$ & $3$ & $3$\\
\hline
\end{tabular}
\end{center}
}


\begin{thebibliography}{99}

\bibitem{Aad:2012tfa}
  G.~Aad {\it et al.}  [ATLAS Collaboration],
  ``Observation of a new particle in the search for the Standard Model Higgs boson with the ATLAS detector at the LHC,''
  Phys.\ Lett.\ B {\bf 716} (2012) 1
  [arXiv:1207.7214 [hep-ex]].

\bibitem{Chatrchyan:2012ufa}
  S.~Chatrchyan {\it et al.}  [CMS Collaboration],
  ``Observation of a new boson at a mass of 125 GeV with the CMS experiment at the LHC,''
  Phys.\ Lett.\ B {\bf 716} (2012) 30
  [arXiv:1207.7235 [hep-ex]].

\bibitem{Arbey:2011ab}
  A.~Arbey, M.~Battaglia, A.~Djouadi, F.~Mahmoudi and J.~Quevillon,
  ``Implications of a 125 GeV Higgs for supersymmetric models,''
  Phys.\ Lett.\ B {\bf 708} (2012) 162
  [arXiv:1112.3028 [hep-ph]].

\bibitem{Draper:2011aa}
  P.~Draper, P.~Meade, M.~Reece and D.~Shih,
  ``Implications of a 125 GeV Higgs for the MSSM and Low-Scale SUSY Breaking,''
  Phys.\ Rev.\ D {\bf 85} (2012) 095007
  [arXiv:1112.3068 [hep-ph]].  
    
\bibitem{Dine:1996xk}
  M.~Dine, Y.~Nir, Y.~Shirman,
  ``Variations on minimal gauge mediated supersymmetry breaking,''
  Phys.\ Rev.\  {\bf D55 } (1997)  1501-1508.
  [hep-ph/9607397].  
    
\bibitem{Giudice:1997ni}
  G.~F.~Giudice, R.~Rattazzi,
  ``Extracting supersymmetry breaking effects from wave function renormalization,''
  Nucl.\ Phys.\  {\bf B511 } (1998)  25-44.
  [hep-ph/9706540].

\bibitem{Chacko:2001km}
  Z.~Chacko, E.~Ponton,
  ``Yukawa deflected gauge mediation,''
  Phys.\ Rev.\  {\bf D66 } (2002)  095004.
  [hep-ph/0112190].

\bibitem{Evans:2013kxa}
  J.~A.~Evans and D.~Shih,
  ``Surveying Extended GMSB Models with mh=125 GeV,''
  arXiv:1303.0228 [hep-ph].

\bibitem{Giudice:1998bp}
  G.~F.~Giudice, R.~Rattazzi,
  ``Theories with gauge mediated supersymmetry breaking,''
  Phys.\ Rept.\  {\bf 322 } (1999)  419-499.
  [hep-ph/9801271].      
    

\bibitem{Kang:2012ra}
  Z.~Kang, T.~Li, T.~Liu, C.~Tong and J.~M.~Yang,
  ``A Heavy SM-like Higgs and a Light Stop from Yukawa-Deflected Gauge Mediation,''
  Phys.\ Rev.\ D {\bf 86} (2012) 095020
  [arXiv:1203.2336 [hep-ph]].  

\bibitem{Craig:2012xp}
  N.~Craig, S.~Knapen, D.~Shih and Y.~Zhao,
  ``A Complete Model of Low-Scale Gauge Mediation,''
  JHEP {\bf 1303} (2013) 154
  [arXiv:1206.4086 [hep-ph]].

\bibitem{Evans:2011be}
  J.~L.~Evans, M.~Ibe, T.~T.~Yanagida,
  ``Relatively Heavy Higgs Boson in More Generic Gauge Mediation,'' 
  [arXiv:1107.3006 [hep-ph]].

\bibitem{Shadmi:2011hs}
  Y.~Shadmi, P.~Z.~Szabo,
  ``Flavored Gauge-Mediation,''
  [arXiv:1103.0292 [hep-ph]].
  
\bibitem{Albaid:2012qk}
  A.~Albaid and K.~S.~Babu,
  ``Higgs boson of mass 125 GeV in GMSB models with messenger-matter mixing,''
  arXiv:1207.1014 [hep-ph].  
 
\bibitem{Abdullah:2012tq}
  M.~Abdullah, I.~Galon, Y.~Shadmi and Y.~Shirman,
  ``Flavored Gauge Mediation, A Heavy Higgs, and Supersymmetric Alignment,''
  arXiv:1209.4904 [hep-ph]. 

\bibitem{Calibbi:2013mka}
  L.~Calibbi, P.~Paradisi and R.~Ziegler,
  ``Gauge Mediation beyond Minimal Flavor Violation,''
  arXiv:1304.1453 [hep-ph].  
    
\bibitem{Byakti:2013ti}
  P.~Byakti and T.~S.~Ray,
  ``Burgeoning the Higgs mass to 125 GeV through messenger-matter interactions in GMSB models,''
  arXiv:1301.7605 [hep-ph].  
  
\bibitem{Kats:2011it}
  Y.~Kats, D.~Shih,
  ``Light Stop NLSPs at the Tevatron and LHC,''
  JHEP {\bf 1108 } (2011)  049.
  [arXiv:1106.0030 [hep-ph]].

\bibitem{Dvali:1996cu}
  G.~R.~Dvali, G.~F.~Giudice and A.~Pomarol,
  ``The Mu problem in theories with gauge mediated supersymmetry breaking,''
  Nucl.\ Phys.\ B {\bf 478} (1996) 31
  [hep-ph/9603238].  
      
\bibitem{Jelinski:2011xe}
  T.~Jelinski, J.~Pawelczyk and K.~Turzynski,
  ``On Low-Energy Predictions of Unification Models Inspired by F-theory,''
  Phys.\ Lett.\ B {\bf 711} (2012) 307
  [arXiv:1111.6492 [hep-ph]].    

\bibitem{Barbier:2004ez}
  R.~Barbier, C.~Berat, M.~Besancon, M.~Chemtob, A.~Deandrea, E.~Dudas, P.~Fayet and S.~Lavignac {\it et al.},
  ``R-parity violating supersymmetry,''
  Phys.\ Rept.\  {\bf 420} (2005) 1
  [hep-ph/0406039].

\bibitem{primer}
  S.~P.~Martin,
  ``A Supersymmetry primer,''
  In *Kane, G.L. (ed.): Perspectives on supersymmetry* 1-98.
  [arXiv:hep-ph/9709356 [hep-ph]].

\bibitem{Dudas:2010zb}
  E.~Dudas and E.~Palti,
  ``On hypercharge flux and exotics in F-theory GUTs,''
  JHEP {\bf 1009} (2010) 013
  [arXiv:1007.1297 [hep-ph]].

\bibitem{Buchmuller:2004eg}
  W.~Buchmuller, L.~Covi, D.~Emmanuel-Costa and S.~Wiesenfeldt,
  ``Flavour structure and proton decay in 6D orbifold GUTs,''
  JHEP {\bf 0409} (2004) 004
  [hep-ph/0407070].

\bibitem{Heckman:2009mn}
  J.~J.~Heckman, A.~Tavanfar, C.~Vafa,
  ``The Point of E(8) in F-theory GUTs,''
  JHEP {\bf 1008 } (2010)  040.
  [arXiv:0906.0581 [hep-th]].

\bibitem{Pawelczyk:2010xh}
  J.~Pawe{\l}czyk,
  ``F-theory inspired GUTs with extra charged matter,''
  Phys.\ Lett.\  {\bf B697 } (2011)  75-79.
  [arXiv:1008.2254 [hep-ph]].
  
\bibitem{Dolan:2011aq}
  M.~J.~Dolan, J.~Marsano and S.~Schafer-Nameki,
  ``Unification and Phenomenology of F-Theory GUTs with $U(1)_{PQ},$''
  [arXiv:1109.4958 [hep-ph]].

\bibitem{Djouadi:2002ze}
  A.~Djouadi, J.~-L.~Kneur, G.~Moultaka,
  ``SuSpect: A Fortran code for the supersymmetric and Higgs particle spectrum in the MSSM,''
  Comput.\ Phys.\ Commun.\  {\bf 176 } (2007)  426-455.
  [hep-ph/0211331].    
  
\bibitem{Aad:2012fqa}
  G.~Aad {\it et al.}  [ATLAS Collaboration],
  ``Search for squarks and gluinos with the ATLAS detector in final states with jets and missing transverse momentum using 4.7 fb$^{-1}$ of $\sqrt{s}=7$ TeV proton-proton collision data,''
  Phys.\ Rev.\ D {\bf 87} (2013) 012008
  [arXiv:1208.0949 [hep-ex]].

 
\end{thebibliography}
\end{document}